\documentclass[11pt]{article}
\usepackage{amsmath}
\usepackage{epsf,afterpage,epsfig}

\def\abstracts#1#2#3{{
        \centering{\begin{minipage}{4.62in}\baselineskip=13pt
        \small
        \centerline{\bf Abstract}
        \vspace*{0.2cm}
        \parindent=0pt #1\par
        \parindent=18pt #2\par
        \parindent=15pt #3
        \end{minipage} }\par}}
\begin{document}
\vspace*{-2cm}
\hfill \parbox{4cm}{ ~\\~}\\
%\hfill \parbox{5.5cm}{
%                     Leipzig preprint NTZ ??/1999}\\
%\vspace*{2cm}
%
\centerline{\LARGE \bf Non-Self Averaging in Autocorrelations }\\[0.3cm]
\centerline{\LARGE \bf for Potts Models on }\\[0.3cm]
\centerline{\LARGE \bf Quenched Random Gravity Graphs}\\[0.4cm]
%\addtocounter{footnote}{-1}
%\renewcommand{\thefootnote}{\arabic{footnote}}
\vspace*{0.2cm}
\centerline{\large {\em Wolfhard Janke$^1$ and
Desmond A. Johnston$^2$\/}}\\[0.4cm]
\centerline{\large {\small $^1$ Institut f\"ur Theoretische Physik,
                    Universit\"at Leipzig,}}
\centerline{ {\small Augustusplatz 10/11, D-04109 Leipzig, Germany }}\\[0.5cm]
\centerline{\large {\small $^2$ Department of Mathematics,
                    Heriot-Watt University,}}
\centerline{    {\small Edinburgh, EH14\,4AS, Scotland  }}\\[0.5cm]
\abstracts{}{We investigate the non-self-averaging properties
of the dynamics of Ising, 4-state Potts and 
10-state Potts models in single-cluster Monte Carlo simulations
on quenched ensembles
of planar, trivalent ($\Phi^3$) random graphs,
which we use as an example of relevant quenched {\em connectivity\/} disorder.
\\ 
We employ a novel application of scaling techniques to the cumulative 
probability distribution of the autocorrelation times for both the energy and 
magnetisation in order to discern non-self-averaging.
Although the specific results discussed here are for
quenched random graphs, the method has quite general applicability.
}{}
%\vspace*{1.5cm}
%\noindent PACS numbers: 05.50.+q, 75.10.Hk, 64.60.Cn
%
\thispagestyle{empty}
\newpage
\pagenumbering{arabic}

\section{Introduction}

The effect of 
quenched, typically bond, disorder on the critical behaviour of spin systems
has been a subject of interest for many years \cite{00}, both because
of its own intrinsic interest and the prevalence of disordered systems in 
nature. The Harris criterion states \cite{harris} that the critical behaviour
of a pure system will be unchanged 
by the introduction of weak quenched bond disorder if the specific-heat 
exponent of the pure system, $\alpha_p$, is less than zero.
By the same token if $\alpha_p$ is greater than zero
the disordered system will not be governed by the pure fixed point, but rather
a new disordered fixed point. The borderline, 
$\alpha_p=0$, constitutes a marginal case which requires more careful 
investigation for each specific case. 

As was emphasised in \cite{ah1}, a pure fixed point is 
usually characterised by a gaussian 
distribution of renormalised couplings around the fixed point Hamiltonian 
which tends to a delta function 
in the thermodynamic limit, whereas a disordered fixed point might be
expected to be characterised by some other 
distribution which tended to a finite width in the thermodynamic limit
\footnote{As we note in the sequel, the presence of such a 
finite width distribution, and hence non-self-averaging,
is not a hard and fast indicator of new, random fixed points since
Poisonnian random lattices still 
display pure critical behaviour even though they 
also display non-self-averaging
properties.}. For a fixed point
with a finite width distribution of couplings a measurement of the density of
an extensive thermodynamic quantity such as the energy $E$, magnetisation $M$,
specific heat $C$ or magnetic susceptibility $\chi$, would be different on 
each sample because of the different disorder realizations. Such behaviour is
called non-self-averaging.

This non-self-averaging behaviour for an observable
$X$ (such as $E$, $M$, $C$, or $\chi$ above) can be characterised by examining
the normalised variance
\begin{equation}
R ( X ) = { [X^2 ]_{\rm av} - [X ]_{\rm av}^2 \over [X ]_{\rm av} ^2},
\label{eq:R}
\end{equation}
where $[\dots]_{\rm av}$ denotes an average over the disorder realizations
\cite{ah1,wd1,wd2}.
If $R ( X ) \to 0$ as $N \to \infty$, where $N$ is the number of lattice sites,
then we have self-averaging, whereas if $R(X) \to c$, 
with $c$ a constant, we have non-self-averaging. 
Self-averaging systems may in turn be divided into strongly self-averaging 
systems where $R ( X ) \sim N^{-1}$, which is the typical behaviour
off criticality and at pure fixed points and weakly self-averaging systems 
where $R ( X )
\sim  N^{\kappa}$ with $ -1 < \kappa <0$, which has been observed for 
the case of irrelevant quenched disorder at criticality \cite{wd1}.
For self-averaging
systems measurements on a single large system are sufficient, whereas
for non-self-averaging systems measurements on different realizations of
the disorder {\it must} be carried out in order 
to obtain reliable ensemble averages.

In this paper we will be concerned with quenched connectivity
disorder, which has received rather less attention
than quenched bond disorder. 
The possible influence of quenched geometrical disorder 
(connectivity, aperiodicity, \dots)  
on the universality properties of statistical mechanical 
systems in general has been explored by Luck \cite{luck}
who arrived at a criterion rather similar in spirit to the Harris criterion
for the (ir)relevance of such disorder.
He noted that if $B ( \Omega )$ was the number of bonds in a region $\Omega$
and $\Sigma ( \Omega) = \sum_{<ij> \epsilon \Omega} J_{i,j}$ was the sum
of bond values in that region then, although both these quantities scaled
as the volume $N$ of the region $\Omega$, one had
\begin{equation}
\Sigma (\Omega) - J_0 B ( \Omega) \sim N^{\phi},
\end{equation}
where $J_0$ was the limiting value as $N \to \infty$ of 
$\Sigma ( \Omega) / B(\Omega)$ and $0 \le \phi < 1$ was a fluctuation 
(or wandering) exponent.
The geometrical fluctuations were found to be relevant if
\begin{equation}
\phi >  { 1 - \alpha \over 2 -\alpha}.
\end{equation}

Although explicit calculations of $\phi$ have not been 
carried out for models with quenched connectivity disorder, 
simulations have shown that it appears 
to be remarkably difficult to escape from 
pure fixed points with such disorder, in contrast to the case of
bond disorder. 
A prime example of this generic behaviour
is Poisonnian (or Voronoi) random lattices,
which have been shown to display 
the pure critical exponents for the 3D 
Ising model \cite{janke_vill} to a very high degree
of accuracy. Numerous other systems with quenched connectivity
disorder show similar, pure critical behaviour \cite{jv}.
However, the quenched connectivity disorder manifested
by an ensemble of planar, trivalent  random 
graphs, which we denote $\Phi^3$ graphs for brevity, 
{\it does} appear to give rise to new disordered fixed points
\cite{janke_des2}.
The exponents of the Ising and $q \le 4$ state Potts
models, which already possess continuous transitions on flat $2D$
lattices, are altered 
and the first-order transition of higher state Potts
models is softened to a continuous transition \cite{janke_des2,janke_des1} 
on the $\Phi^3$ graphs.

The $\Phi^3$ graphs in question are precisely those generated in a simulation 
of pure 2D quantum gravity, though in that case one has an
annealed ensemble in which the connectivities
fluctuate.
For such an annealed ensemble of $\Phi^3$ graphs 
the KPZ formula \cite{KPZ} shows how the conformal weights of operators in 
spin models living on the graphs
are transformed from their flat $2D$ lattice values.
The KPZ formula {\it per se} thus applies to systems in which the
spins are fluctuating on the same time-scale as the connectivity.
One can, however, obtain predictions for
the critical exponents of spin models living on an quenched ensemble 
of such graphs \footnote{In which the connectivity disorder is frozen in.}, 
which is what we are considering here,
by taking a quenched limit in the KPZ formula \cite{cardy,des,wrong}.
In this limit one finds that
the flat lattice conformal weight $\Delta$ is transmuted to
a new quenched weight 
% $\tilde \Delta_{quenched}$
\begin{equation}
\tilde \Delta_{quenched} = { \sqrt{1 + 24 \Delta} - {1 } \over 4 },
\label{eq:oops}
\end{equation}                     
which may then be used to determine the critical exponents $\alpha$ and $\beta$
for the quenched ensemble\footnote{
It is only fair to remark that the analysis of the data
in \cite{janke_des2}, on which the current paper is also based, only lend 
rather weak support to these predictions.}.

To date investigations of non-self-averaging behaviour in the presence of 
quenched disorder have concentrated exclusively on static quantities such as 
the susceptibility, rather than the dynamical properties described
by the autocorrelation times for various observables.
In this paper we study the non-self-averaging properties
of autocorrelation times 
for the energy and the magnetisation in the presence of quenched 
(in our case connectivity) disorder. We
consider the Ising, $q=4$ and $q=10$ state Potts models on 
quenched ensembles of
$\Phi^3$ random graphs, since the simulations of
\cite{janke_des2}
provided strong evidence 
for new, disordered fixed points in all these cases.

In the next section we briefy recall the models studied and describe the
Monte Carlo simulations performed. The results of the autocorrelation analyses
are presented in Sec.~3, and in Sec.~4 we close with a summary and a few
concluding remarks.

\section{The Model and Simulations}

As in \cite{janke_des2}
we use the standard definition of the $q$-state Potts model
partition function and energy
\begin{equation}
Z_{\rm Potts} = \sum_{\{\sigma_i\}} e^{-\beta E}; \; \;
E = -\sum_{\langle ij \rangle}
\delta_{\sigma_i \sigma_j};  \; \; \sigma_i = 1,\dots,q,
\label{eq:zpotts}
\end{equation}   
where $\beta=J/k_B T$ is the inverse temperature in natural units,
$\delta$ is the Kronecker symbol, and
$\langle ij \rangle$ denotes the nearest-neighbour bonds of the random
$\Phi^3$ graphs (without tadpoles or self-energy bubbles) with
$N$ sites. We consider the cases $q=2$ and $4$
with $N = 500$,
1\,000, 2\,000, 3\,000, 4\,000, 5\,000, and 10\,000 which in the
pure model exhibit second-order phase transitions, and the case
$q=10$ with $N = 250$, 500, 1\,000, 2\,000, 3\,000, 5\,000,
and 10\,000 which in the pure model undergoes a first-order
phase transition.               

The simulations were carried out using the Wolff single-cluster update
algorithm \cite{13}.
For each
lattice size we generated 64 independent graphs using the Tutte algorithm
\cite{14}, and performed 500K equilibration sweeps followed by
up to 10 million measurement sweeps  in order to obtain 500K independent
measurement sweeps for each lattice size.
The runs were carried out at several
$\beta$ values near the transition point
and time series of the energy $E$ and the magnetisation%
\footnote{Where $n_i \le N$ denotes the number of spins of ``orientation''
 $i=1,\dots,q$ in one lattice configuration.}
$M = (q \, {\rm max} \{ n_i \} - N)/(q-1)$
recorded for each graph.
In what follows the per-site quantities are denoted by $e = E/N$ and $m = M/N$,
the thermal averages on each individual graph by $\langle \ldots \rangle$ 
and the quenched average over the different graphs by $[ \ldots ]_{\rm av}$.
From the time series of $e$ and $m$ it is straightforward to
compute in the finite-size scaling (FSS) region various quantities at
nearby values of $\beta$  by
standard re-weighting \cite{15} techniques.                          

To estimate the statistical (thermal) errors for each of the 64 realizations,
the time-series data was split into
bins, which were jack-knifed \cite{16} to decrease the bias in the
analysis of re-weighted data.
The final values are averages over the 64 realizations which will be denoted
by square brackets $[\dots]_{\rm av}$, and the error bars are computed from the
fluctuations among the realizations. Note that these errors contain both
the average thermal error for a given realization and the theoretical variance
for infinitely accurate thermal averages which is caused by the variation
over the random graphs.

From the time series of the energy measurements we computed by re-weighting the
average energy, specific heat, and energetic fourth-order cumulant,
as discussed in more detail in \cite{janke_des2}.
Similarly, we derived from the magnetisation measurements the average
magnetisation, susceptibility, and magnetic cumulants and also
evaluated
mixed quantities involving both the energy and magnetisation. 
However, it is the dynamical aspects of the simulations 
which are our principal concern here and these
are characterised by the autocorrelation
functions and the associated integrated autocorrelation times $\hat{\tau}$.
It is now customary when discussing single-cluster algorithm 
simulations \cite{1c}
to convert the $\hat{\tau}$ thus obtained by multiplying with a factor
$f = n_{\rm flip} \langle |C| \rangle/N$ to a 
standardised scale where, on the average,
measurements are taken after every spin has been flipped once.
This allows a fair comparison with,
e.g., Metropolis simulations. 

When one has quenched random disorder this procedure
is not unique due to the average over realizations ($[ ... ]_{av}$), 
since one can take either  $[\tau]_{\rm av} \equiv
[f \cdot \hat{\tau}]_{\rm av}$ or $[f]_{\rm av} \cdot [\hat{\tau}]_{\rm av}$.
We have presented the raw data and both variations in Tables 1--6
for both the energy and magnetisation for $q=2$, $4$, and $10$ where it can 
be seen that the differences between the
two averaging prescriptions are rather small,
so for all practical purposes they can be considered to be equivalent.
For definiteness in the scaling analysis we take 
$[\tau]_{\rm av} \equiv [f \cdot \hat{\tau}]_{\rm av}$. In Tables 1--6
the minimum value of $\tau \;  
( \; = f \cdot \hat{\tau})$ for the various realizations 
is denoted by $\tau_{\rm min}$ and the maximum by $\tau_{\rm max}$. The standard
deviation $\Delta \tau$ and its scaled form, $\Delta \tau / [ \tau ]_{\rm av}$, are also tabulated.

\section{Results for Autocorrelation Times}

The integrated autocorrelation times for each random-graph realization
are obtained by blocking techniques. These are (necessarily) measurements 
at the simulation points which are chosen  close to a 
finite-size scaling sequence of $\beta$-values but {\em not\/} with high 
precision since these are {\em a priori\/} estimates based on the results on
smaller lattices. For higher accuracy one would have to redo the simulations 
using the knowledge of the infinite-volume estimates of $\beta_c$ or the 
locations of the maxima of $C$, $\chi$, etc. obtained from the present batch 
of runs and presented in \cite{janke_des2}. 
As a consequence of this choice of measurement points it can be seen in the 
sequel that one or two data points lie rather far from the general trend.

Looking at the behaviour of the autocorrelation times for each $q$ in turn, 
we can see that
the autocorrelation times for $q=2$ stay roughly constant with increasing
system size for both the energy and the magnetisation. For the energy we 
obtain values in the range $3-4$, and for the magnetisation in the range
$1.6-2.2$. These results are obtained with the cluster update for 
{\em general\/} $q$-state Potts models where the new spin direction 
$s_{\rm new}$ for a cluster is chosen randomly from $s_{\rm new} \in [1,q]$. 
Picking the old value, $s_{\rm new} = s_{\rm old}$, would not change anything.
Hence for $q=2$ the autocorrelation time can be reduced by a factor of 2 by
requiring that the spin direction of the cluster flips rather than is chosen
randomly. 

For $q=4$ the autocorrelation times are also still reasonably small,
covering a range $12-18$ for the energy and a range $7-10$ for the
magnetisation. However, here a scaling with system size is now clearly 
observable and fits to the standard finite-size scaling ansatz

\begin{equation}
[\tau]_{\rm av} = a N^{z/D}
\end{equation}
give for the energy $\ln a = 2.283(82)$ and $z/D = 0.064(10)$ with
$\chi^2/{\rm dof} = 1.28$ or a goodness-of-fit parameter $Q=0.28$, if 
the $N=500$ and $N=1000$ graphs are omitted. For the magnetisation the
fit through all available graph sizes yields
$\ln a = 1.554(95)$ and $z/D = 0.074(13)$ with
$\chi^2/{\rm dof} = 1.01$ or $Q=0.41$.
The data along with the fits are shown in Fig.~1.
Note that the estimates of $z/D$ on the random graphs are considerably 
smaller than for regular lattices 
\cite{17} where
$z/D = 0.876(11)/2 = 0.438(6)$ for the integrated autocorrelation time of
the energy, using the Swendsen-Wang cluster-update~\footnote{ 
In two dimensions the difference between the dynamical critical exponent
of the Wolff single-cluster and Swendsen-Wang cluster algorithm, if any,
is empirically extremely small.}.
In fact, since
the Li-Sokal bound \cite{18}
guarantees for regular lattices that 
$\tau_e \ge {\rm const} \times C$ and the specific heat 
$C$ diverges on regular lattices at criticality like $L (\ln L)^{-3/2}$, 
we see that (as was noted in \cite{17}) this 
actually must be an
underestimate. It should be emphasised that for quenched (or annealed) 
gravity graphs the singularity of the specific heat is predicted to be 
weakened; hence we are not a priori in contradiction with the suitably 
generalised Li-Sokal bound.

For $q=10$, where one would expect on regular lattices for 
both autocorrelation times a pronounced
exponential increase with system size due to the
first-order nature of the transition, the first point to note is that
here on $\Phi^3$ graphs the values are very large by comparison with the Ising
and $q=4$ Potts model measurements discussed above, falling into the range 
$60-500$ for the energy and $40 - 350$ for the magnetisation. The increase of 
$[\tau]_{\rm av}$ with system size, 
however, is consistent with a power-law scaling behaviour rather than the 
exponential increase of a first-order transition.
If we omit the two smallest graph sizes with $N=250$ and
$N=500$ we obtain from fits of the energy autocorrelations the estimates 
$\ln a = -1.44(29)$ and $z/D = 0.829(35)$ with
$\chi^2/{\rm dof} = 1.97$ or $Q=0.12$.
Fits of the magnetisation autocorrelations yield
$\ln a = -3.49(44)$ and $z/D = 1.019(54)$ with     
$\chi^2/{\rm dof} = 1.00$ or $Q=0.39$. The data and the quality of the
fits can be inspected in Fig.~2.

For quenched, random systems not only the scaling behaviour of the
average $[\tau]_{\rm av}$ is of interest but also the properties of 
the whole probability density $P(\tau)$. Since we have only few ($ = 64$)
events for sampling this density it is numerically (and mathematically)
more sensible to consider the cumulative probability distribution
$F(\tau) = \int_0^\tau P(\tau') d\tau'$, with the obvious relation
$d F(\tau)/d\tau = P(\tau)$. 

In Figs.~3 and 4 we have plotted the cumulative distribution of the 
scaled $\tau$ at the simulation point in the
$q=4$ model for both the energy and the magnetisation.
It is clear from the graphs
that the distribution is slightly broadening with increasing lattice size, 
rather than sharpening.
For self-averaging measurements one would expect the curves to
tend to a step function,
since the underlying probability density
of $\tau$ would tend to a delta-function in such a case. One can therefore
conclude that the measurements
of $\tau$ in the simulations provide strong evidence for 
{\it non}-self-averaging
behaviour in the $q=4$ model. This statement can be made more quantitative
by considering similar to eq.~(\ref{eq:R}) the ratio of the width of the 
probability density, the standard deviation $\Delta \tau$, and the average 
value, $[\tau]_{\rm av}$, which sets the scale.
As this is a property of the
quenched randomness here we tacitly assume that the thermal noise of the
estimates of $\tau$ for a given realization can be neglected which, in view
of our extremely high statistics, is justified.

The plot of 
$\Delta \tau/[\tau]_{\rm av}$ in Fig.~5 shows that the relative
widths of the densities of both the energy and the magnetisation stay roughly 
constant with increasing system size, thus clearly 
demonstrating the lack of self-averaging. Another way to demonstrate this
property graphically is to plot the probability distribution $F$ against the
scaled variable $\tau/[\tau]_{\rm av}$. If the density is non-self-averaging
with $\Delta \tau/[\tau]_{\rm av} = {\rm const}$ one expects 
to see in such a plot data collapse onto a single master-curve. As can be 
seen in Figs.~6 and 7
this is indeed the case for both the energy and the magnetisation.

For $q=2$ the corresponding plots look very similar and are not reproduced here.
More interesting is the qualitatively different case of the $q=10$ Potts model,
because here the first-order
transition on regular lattices is softened to a second-order transition for
quenched, random graphs. Here the plots of the cumulative distributions
in Figs.~\ref{fig8a} and \ref{fig9a} again clearly exhibit the broadening 
with increasing lattice size
just as for $q=4$, rather than 
sharpening. Therefore it is even more impressive to observe that replotting the
data versus the scaled variable $\Delta \tau/[\tau]_{\rm av}$ still produces
well-defined master curves for both the energy and magnetisation
as can be seen in Figs.~\ref{fig8} and \ref{fig9}. The
scaling of $\Delta \tau/[\tau]_{\rm av}$ as a function of system size looks
slightly more scattered than for $q=4$ (recall that the effective statistics
per graph is smaller by about one order of magnitude compared with the $q=4$
model), but also here it is safe to claim that $\Delta \tau/[\tau]_{\rm av}$
stays roughly constant with increasing system size. 
One can therefore conclude that the measurements
of $\tau$ in the simulations provide strong evidence for 
{\it non}-self-averaging behaviour in the $q=10$ model as well.

\section{Conclusions}

Our previous analysis of the static properties of 
simulations of the Ising and $q=4,10$ state Potts models
on $\Phi^3$ graphs
showed that the quenched connectivity disorder
they possessed
altered the exponents of models with a
continuous transition on a regular lattice, and softened
the first-order transition of the $q=10$ model to a continuous transition.

The analysis of the autocorrelation times discussed here shows that these
models display another property that is often associated with a 
disordered fixed point with a distribution of couplings, namely, 
non-self-averaging. One must, however, be a little careful in using
non-self-averaging as a diagnostic for distinguishing pure and random 
fixed points:
the Ising model on a $3D$ Poisonnian random lattice has been shown to
have the standard critical exponents to a very high degree of accuracy,
but nonetheless still displays non-self-averaging of both static and dynamical
properties \cite{janke_vill}.

Whatever the circumstances in which non-self-averaging appears the 
analysis here shows that it is also manifest in the autocorrelation
times of the systems in question and amenable to a quantitative
scaling analysis. Although we have only discussed the Ising
and $q=4,10$ state Potts models on $\Phi^3$ random graphs here,
it is clear that the idea of looking at  $P(\tau)$ and its
scaling properties to discern non-self-averaging is generally applicable.

%
%---------------------------------------------------------
%
\section*{Acknowledgements}
%
%---------------------------------------------------------
%
DJ was partially supported by a Royal Society of
Edinburgh/SOEID Support Research Fellowship.
WJ acknowledges partial support by the German-Israel-Foundation (GIF) under
contract No.\ I-0438-145.07/95.
The collaborative work of DJ and WJ was funded by ARC grant
313-ARC-XII-98/41. The numerical simulations were performed on a T3D parallel
computer of Zuse-Zentrum f\"ur Informationswissenschaften Berlin (ZIB)
under grant No.~bvpf01.

\addtolength{\textwidth}{1cm}
\clearpage

\begin{table}
\begin{center}
\begin{tabular}{|r|r|r|r|r|r|r|r|r|r|}
\hline
\multicolumn{1}{|c|}{$N$} &
\multicolumn{1}{c|}{$n_{\rm flip}$}  &
\multicolumn{1}{c|}{$[\langle |C| \rangle]$} &
\multicolumn{1}{c|}{$[\hat{\tau}]$} &
\multicolumn{1}{c|}{$[f] [\hat{\tau}]$} &
\multicolumn{1}{c|}{$[\tau] = [f \hat{\tau}]$} &
\multicolumn{1}{c|}{$\tau_{\rm min}$} &
\multicolumn{1}{c|}{$\tau_{\rm max}$} &
\multicolumn{1}{c|}{$\Delta \tau$} &
\multicolumn{1}{c|}{$\Delta \tau/[\tau]$} \\
\hline
  500 & 8 & 172.4(2.8) & 1.18(2) & 3.25 & 3.21(3) & 2.74 & 3.78 &0.2069 &0.0645\\
 1000 &12 & 258.8(4.6) & 1.15(2) & 3.56 & 3.50(3) & 2.79 & 4.04 &0.2193 &0.0626\\
 2000 &20 & 386.0(8.0) & 1.00(2) & 3.87 & 3.79(4) & 3.32 & 4.45 &0.2463 &0.0650\\
 3000 &12 & 896(20)    & 1.09(3) & 3.90 & 3.80(4) & 3.39 & 4.29 &0.2584 &0.0679\\
 4000 & 5 &1076(21)    & 1.39(3) & 1.87 & 1.83(2) & 1.48 & 2.11 &0.1178 &0.0644\\
 5000 & 8 &1305(26)    & 1.83(4) & 3.83 & 3.74(3) & 3.08 & 4.49 &0.2426 &0.0649\\
10000 &10 &2328(53)    & 1.75(5) & 4.07 & 3.94(3) & 3.38 & 4.48 &0.2316 &0.0587\\
\hline
\end{tabular}
\end{center}
\caption{$q=2$: Autocorrelation times of the energy ($f = n_{\rm flip} \langle |C| \rangle /N$).}
\end{table}

\begin{table}
\begin{center}
\begin{tabular}{|r|r|r|r|r|r|r|r|r|r|}
\hline
\multicolumn{1}{|c|}{$N$} &
\multicolumn{1}{c|}{$n_{\rm flip}$}  &
\multicolumn{1}{c|}{$[\langle |C| \rangle]$} &
\multicolumn{1}{c|}{$[\hat{\tau}]$} &
\multicolumn{1}{c|}{$[f] [\hat{\tau}]$} &
\multicolumn{1}{c|}{$[\tau] = [f \hat{\tau}]$} &
\multicolumn{1}{c|}{$\tau_{\rm min}$} &
\multicolumn{1}{c|}{$\tau_{\rm max}$} &
\multicolumn{1}{c|}{$\Delta \tau$} &
\multicolumn{1}{c|}{$\Delta \tau/[\tau]$} \\
\hline
  500 & 8 & 172.4(2.8) & 0.67(1) & 1.85 & 1.85(4) & 1.17 & 2.48 &0.2888 &0.1559\\
 1000 &12 & 258.8(4.6) & 0.61(1) & 1.91 & 1.91(4) & 1.23 & 2.63 &0.2977 &0.1561\\
 2000 &20 & 386.0(8.0) & 0.55(1) & 2.14 & 2.14(5) & 1.47 & 3.39 &0.3895 &0.1821\\
 3000 &12 & 896(20)    & 0.61(1) & 2.19 & 2.19(6) & 1.18 & 3.00 &0.4078 &0.1859\\
 4000 & 5 &1076(21)    & 0.66(1) & 0.89 & 0.89(2) & 0.56 & 1.27 &0.1567 &0.1760\\
 5000 & 8 &1305(26)    & 0.77(1) & 1.61 & 1.62(5) & 0.72 & 2.25 &0.3312 &0.2043\\
10000 &10 &2328(53)    & 0.73(1) & 1.70 & 1.71(5) & 0.79 & 2.45 &0.3693 &0.2159\\
\hline
\end{tabular}
\end{center}
\caption{$q=2$: Autocorrelation times of the magnetisation ($f = n_{\rm flip} \langle |C| \rangle /N$).}
\end{table}

\begin{table}
\begin{center}
\begin{tabular}{|r|r|r|r|r|r|r|r|r|r|}
\hline
\multicolumn{1}{|c|}{$N$} &
\multicolumn{1}{c|}{$n_{\rm flip}$}  &
\multicolumn{1}{c|}{$[\langle |C| \rangle]$} &
\multicolumn{1}{c|}{$[\hat{\tau}]$} &
\multicolumn{1}{c|}{$[f] [\hat{\tau}]$} &
\multicolumn{1}{c|}{$[\tau] = [f \hat{\tau}]$} &
\multicolumn{1}{c|}{$\tau_{\rm min}$} &
\multicolumn{1}{c|}{$\tau_{\rm max}$} &
\multicolumn{1}{c|}{$\Delta \tau$} &
\multicolumn{1}{c|}{$\Delta \tau/[\tau]$} \\
\hline
  500 & 8 & 221.9(4.7) & 3.54(07) & 12.58 & 12.32(14) &  9.77 & 14.67 &1.1122 &0.0903\\
 1000 &12 & 371.1(8.9) & 3.22(07) & 14.32 & 13.89(13) & 11.72 & 16.11 &1.0344 &0.0745\\
 2000 &16 & 639(19)    & 3.24(10) & 16.55 & 15.74(16) & 12.48 & 18.93 &1.2805 &0.0814\\
 3000 &16 & 909(27)    & 3.55(11) & 17.18 & 16.37(21) & 13.30 & 21.06 &1.6273 &0.0994\\
 4000 &18 &1093(28)    & 3.56(11) & 17.54 & 16.83(20) & 12.55 & 20.66 &1.5668 &0.0931\\
 5000 &20 &1634(40)    & 2.71(08) & 17.73 & 16.99(22) & 14.12 & 22.90 &1.6845 &0.0991\\
10000 &20 &3194(81)    & 2.87(11) & 18.36 & 17.37(23) & 13.58 & 22.76 &1.8138 &0.1044\\
\hline
\end{tabular}
\end{center}
\caption{$q=4$: Autocorrelation times of the energy ($f = n_{\rm flip} \langle |C| \rangle /N$).}
\end{table}

\begin{table}
\begin{center}
\begin{tabular}{|r|r|r|r|r|r|r|r|r|r|}
\hline
\multicolumn{1}{|c|}{$N$} &
\multicolumn{1}{c|}{$n_{\rm flip}$}  &
\multicolumn{1}{c|}{$[\langle |C| \rangle]$} &
\multicolumn{1}{c|}{$[\hat{\tau}]$} &
\multicolumn{1}{c|}{$[f] [\hat{\tau}]$} &
\multicolumn{1}{c|}{$[\tau] = [f \hat{\tau}]$} &
\multicolumn{1}{c|}{$\tau_{\rm min}$} &
\multicolumn{1}{c|}{$\tau_{\rm max}$} &
\multicolumn{1}{c|}{$\Delta \tau$} &
\multicolumn{1}{c|}{$\Delta \tau/[\tau]$} \\
\hline

  500 & 8 & 221.9(4.7) & 2.10(4) & 7.47 & 7.50(21) & 3.98 & 10.69 &1.6824 &0.2242\\
 1000 &12 & 371.1(8.9) & 1.74(4) & 7.77 & 7.78(26) & 3.49 & 12.80 &2.0077 &0.2581\\
 2000 &16 & 639(19)    & 1.65(4) & 8.46 & 8.38(27) & 3.52 & 13.11 &2.1223 &0.2534\\
 3000 &16 & 909(27)    & 1.81(5) & 8.79 & 8.80(34) & 3.13 & 16.43 &2.7074 &0.3078\\
 4000 &18 &1093(28)    & 1.70(5) & 8.39 & 8.33(27) & 3.79 & 12.78 &2.1524 &0.2583\\
 5000 &20 &1634(40)    & 1.43(4) & 9.32 & 9.28(30) & 3.48 & 14.03 &2.3661 &0.2549\\
10000 &20 &3194(81)    & 1.45(4) & 9.28 & 9.23(29) & 3.80 & 14.71 &2.2907 &0.2482\\

\hline
\end{tabular}
\end{center}
\caption{$q=4$: Autocorrelation times of the magnetisation ($f = n_{\rm flip} \langle |C| \rangle /N$).}
\end{table}

\begin{table}
\begin{center}
\begin{tabular}{|r|r|r|r|r|r|r|r|r|r|}
\hline
\multicolumn{1}{|c|}{$N$} &
\multicolumn{1}{c|}{$n_{\rm flip}$}  &
\multicolumn{1}{c|}{$[\langle |C| \rangle]$} &
\multicolumn{1}{c|}{$[\hat{\tau}]$} &
\multicolumn{1}{c|}{$[f] [\hat{\tau}]$} &
\multicolumn{1}{c|}{$[\tau] = [f \hat{\tau}]$} &
\multicolumn{1}{c|}{$\tau_{\rm min}$} &
\multicolumn{1}{c|}{$\tau_{\rm max}$} &
\multicolumn{1}{c|}{$\Delta \tau$} &
\multicolumn{1}{c|}{$\Delta \tau/[\tau]$} \\
\hline
   250 &  6 &   67.8(2.1) &  40.3(1.9) & 65.7 &  66.0(4.0) &  23.6 &  176.2 &  31.63 & 0.4796 \\
   500 &  6 &   83.8(3.8) &  74.4(3.2) & 74.9 &  75.5(4.9) &  22.1 &  244.1 &  39.10 & 0.5179 \\
  1000 &  6 &   86.2(3.6) & 134.9(5.4) & 69.8 &  70.2(4.2) &  23.6 &  175.2 &  33.34 & 0.4750 \\
  2000 & 12 &  189(14)    & 111.9(5.2) & 127  & 127(12)    &  34.9 &  555.5 &  90.89 & 0.7163 \\
  3000 & 15 &  362(29)    & 109.5(4.7) & 198  & 185(13)    &  54.6 &  528.1 &  97.01 & 0.5252 \\
  5000 & 15 &  833(52)    & 136.0(6.9) & 340  & 308(18)    &  85.3 &  761.1 & 144.7  & 0.4699 \\
 10000 & 15 & 1652(110)   & 210(11)    & 521  & 452(27)    & 156.0 & 1159.2 & 213.6  & 0.4730 \\
\hline
\end{tabular}
\end{center}
\caption{$q=10$: Autocorrelation times of the energy ($f = n_{\rm flip} \langle |C| \rangle /N$).}
\end{table}

\begin{table}
\begin{center}
\begin{tabular}{|r|r|r|r|r|r|r|r|r|r|}
\hline
\multicolumn{1}{|c|}{$N$} &
\multicolumn{1}{c|}{$n_{\rm flip}$}  &
\multicolumn{1}{c|}{$[\langle |C| \rangle]$} &
\multicolumn{1}{c|}{$[\hat{\tau}]$} &
\multicolumn{1}{c|}{$[f] [\hat{\tau}]$} &
\multicolumn{1}{c|}{$[\tau] = [f \hat{\tau}]$} &
\multicolumn{1}{c|}{$\tau_{\rm min}$} &
\multicolumn{1}{c|}{$\tau_{\rm max}$} &
\multicolumn{1}{c|}{$\Delta \tau$} &
\multicolumn{1}{c|}{$\Delta \tau/[\tau]$} \\
\hline
  250 & 6 &  67.8(2.0) &   24.3(1.2) &   39.6 &   40.5(2.8) &   11.1 & 147.6 & 21.89 & 0.5411\\
  500 & 6 &  83.8(3.8) &   44.2(2.4) &   44.4 &   47.6(4.6) &   10.0 & 251.1 & 36.76 & 0.7715\\
 1000 & 6 &  86.2(3.6) &   58.3(3.4) &   30.2 &   33.0(3.4) &    6.2 & 157.4 & 26.62 & 0.8056\\
 2000 &12 & 189(14)    &   53.1(3.1) &   60.2 &   69.8(9.2) &    8.3 & 472.3 & 73.62 & 1.0540\\
 3000 &15 & 362(29)    &   58.4(3.4) &  105.8 &  108(11)    &   20.1 & 477.5 & 86.48 & 0.7998\\
 5000 &15 & 833(52)    &   82.0(5.6) &  205.1 &  201(17)    &   18.9 & 717.2 &134.44 & 0.6703\\
10000 &15 &1652(110)   &  141.9(8.3) &  351.6 &  335(28)    &   71.9 &1064.4 &222.64 & 0.6639\\
\hline
\end{tabular}
\end{center}
\caption{$q=10$: Autocorrelation times of the magnetisation ($f = n_{\rm flip} \langle |C| \rangle /N$).}
\end{table}

%-----------Figures------------------
\clearpage \newpage
\begin{figure}[htb]
\vskip 20.0truecm
\includegraphics{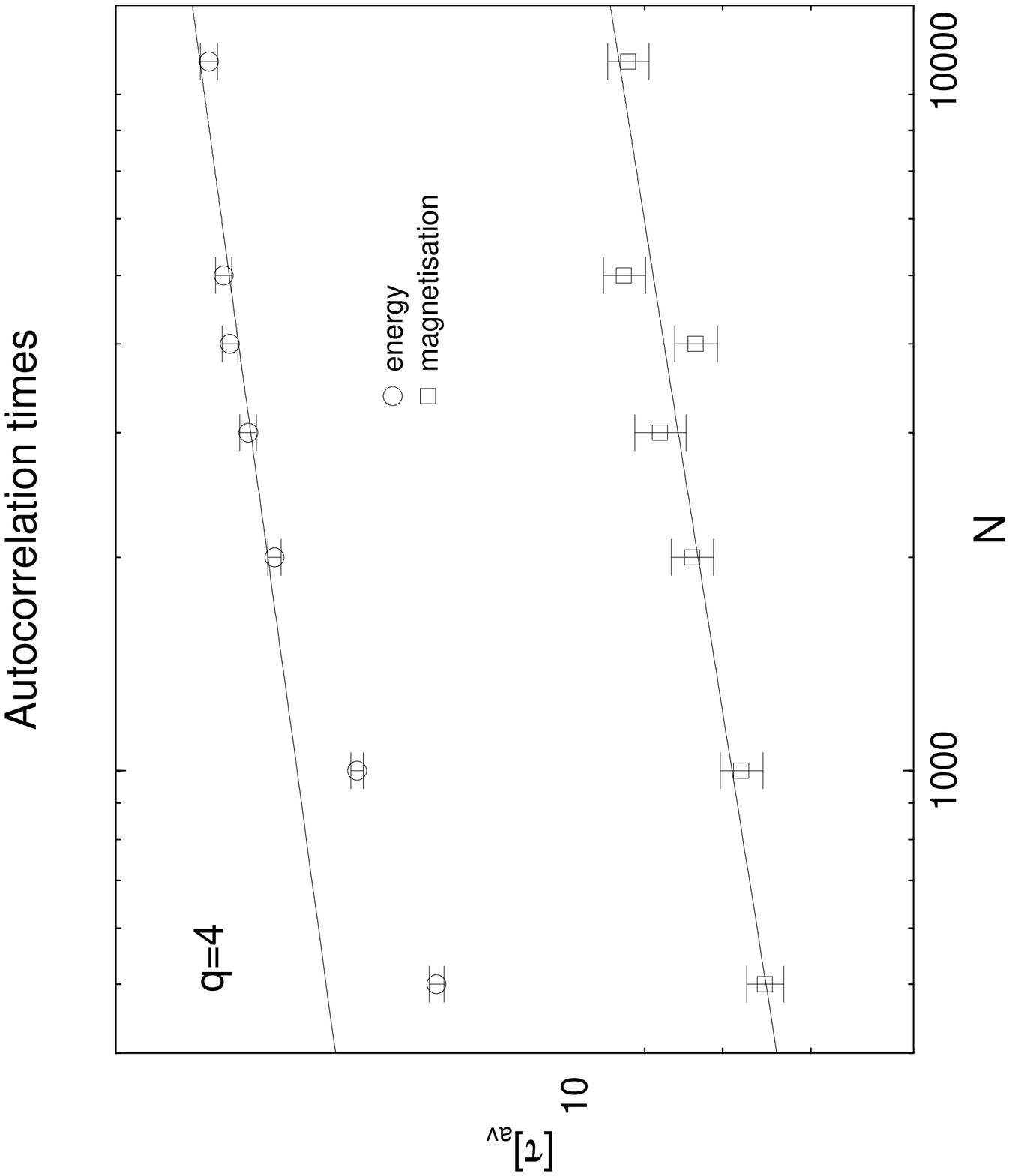}
\caption[]{\label{fig1} The data points and fits to 
$[\tau]_{\rm av} = a N^{z/D}$ for both the energy and magnetisation in 
the $q=4$ state Potts model.
}
\end{figure}                            
%-----------Figures------------------
\clearpage \newpage
\begin{figure}[htb]
\vskip 20.0truecm
\includegraphics{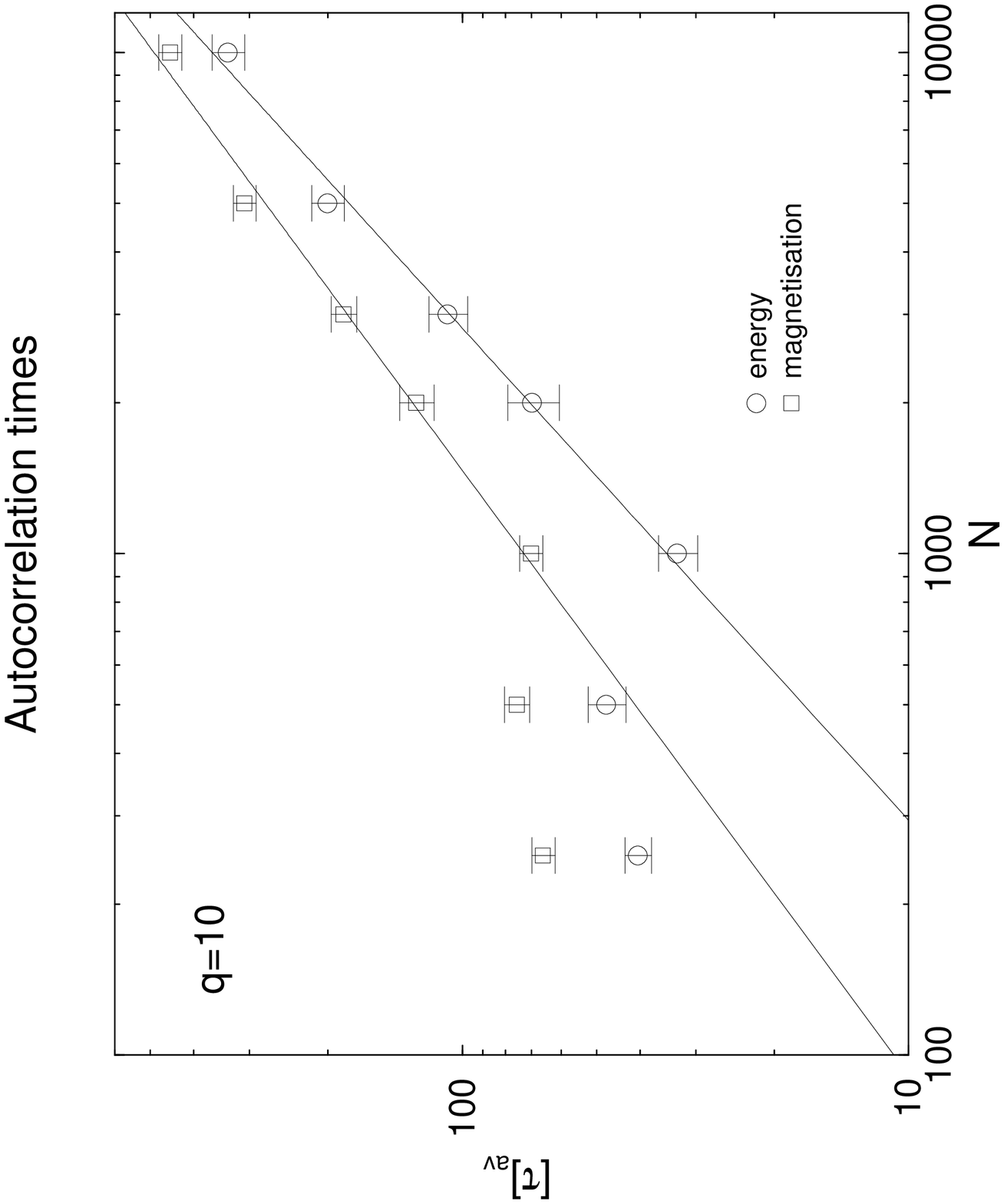}
\caption[]{\label{fig2} The data points and fits to 
$[\tau]_{\rm av} = a N^{z/D}$
for both the energy and magnetisation in the $q=10$ state Potts model.
}
\end{figure}     
%-----------Figures------------------
\clearpage \newpage
\begin{figure}[htb]
\vskip 20.0truecm
\includegraphics{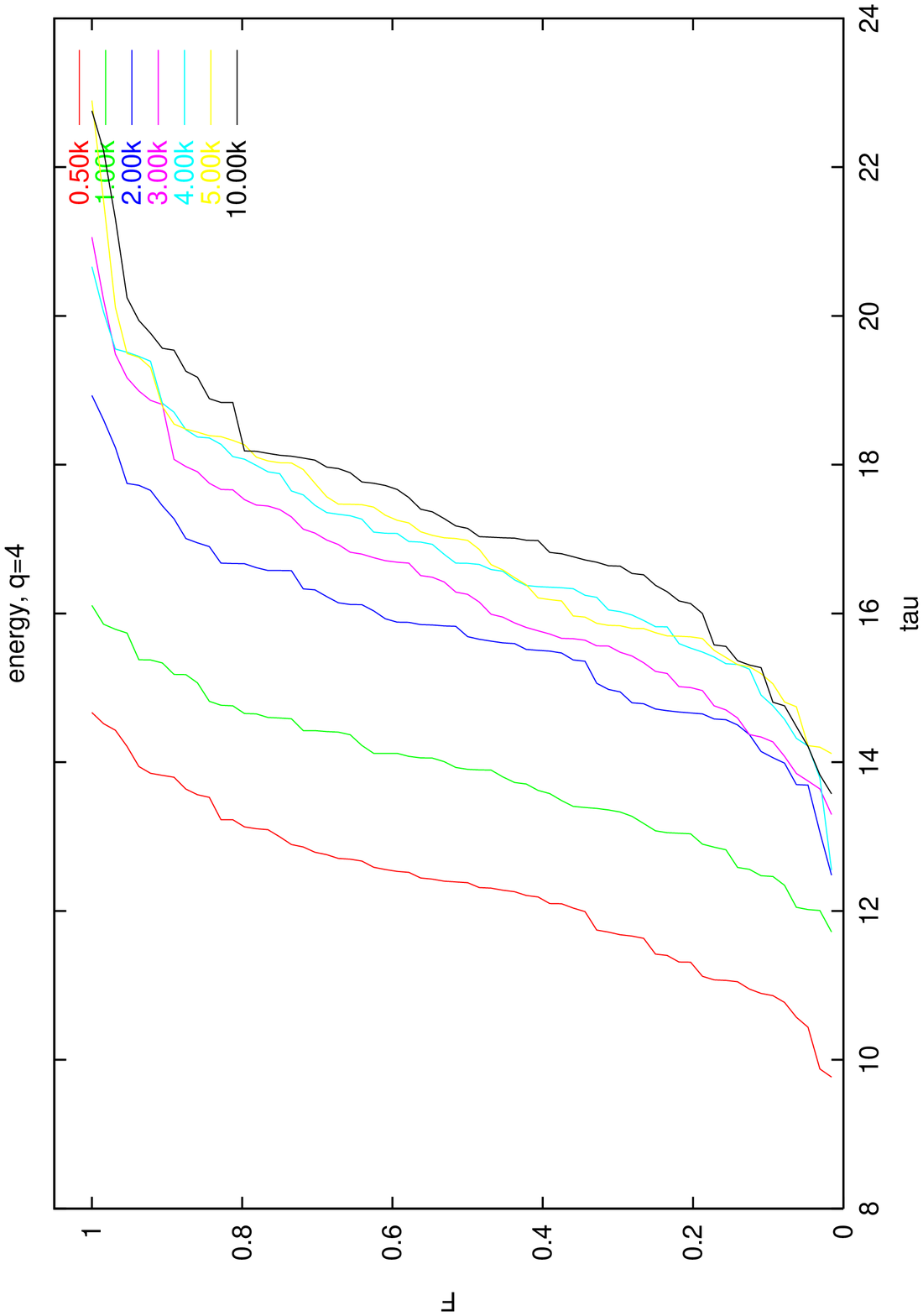}
\caption[]{\label{fig3} The cumulative distribution function for the energy
autocorrelations in the $q=4$ Potts model. 
The system size is {\it increasing} as 
the curves move to the right.
}
\end{figure}                   
%-----------Figures------------------
\clearpage \newpage
\begin{figure}[htb]
\vskip 20.0truecm
\includegraphics{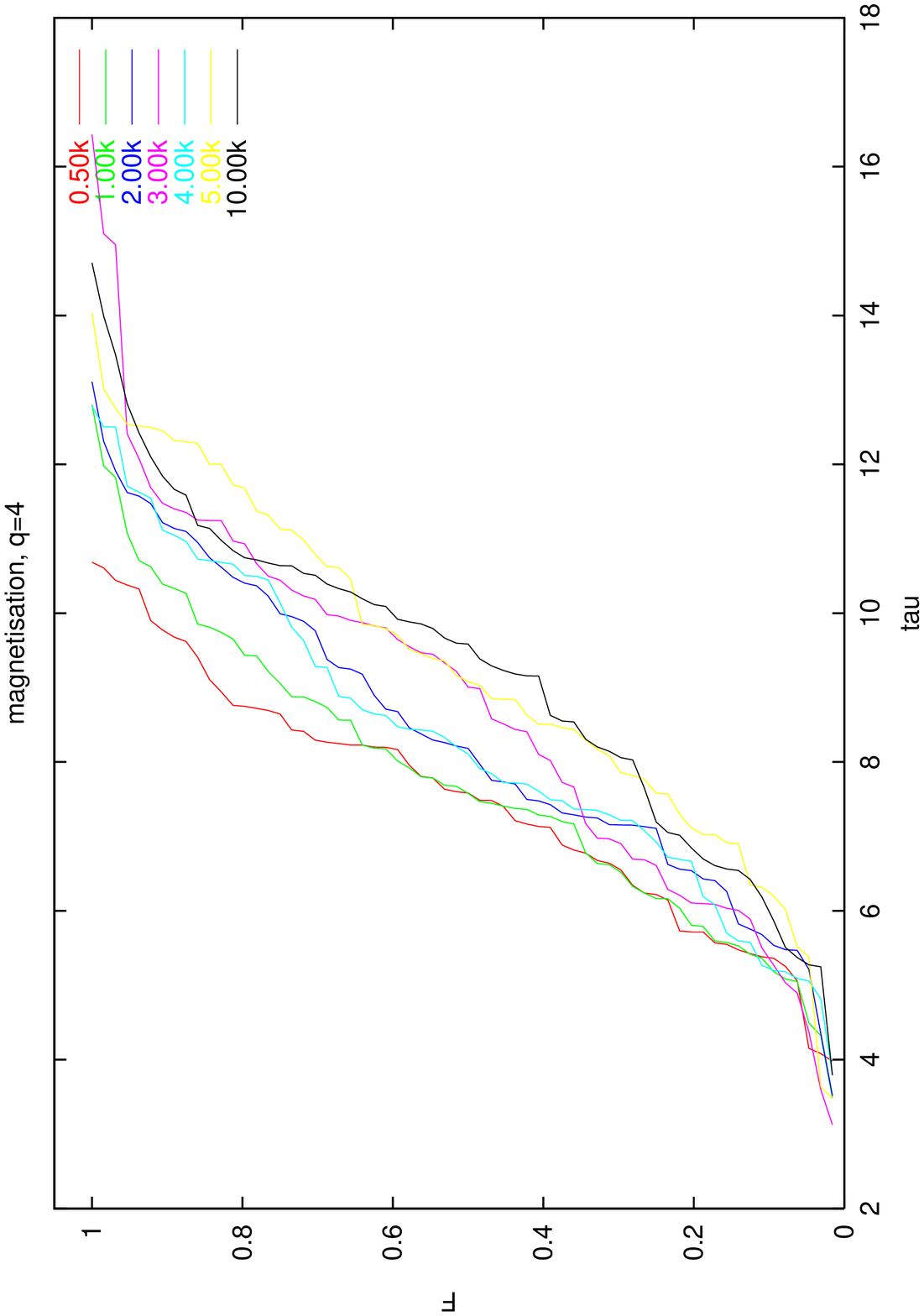}
\caption[]{\label{fig4} The cumulative distribution function 
for the magnetisation autocorrelations in the $q=4$ Potts model.
Again, the system size is increasing as
the curves move to the right.
}
\end{figure}  
%-----------Figures------------------
\clearpage \newpage
\begin{figure}[htb]
\vskip 20.0truecm
\includegraphics{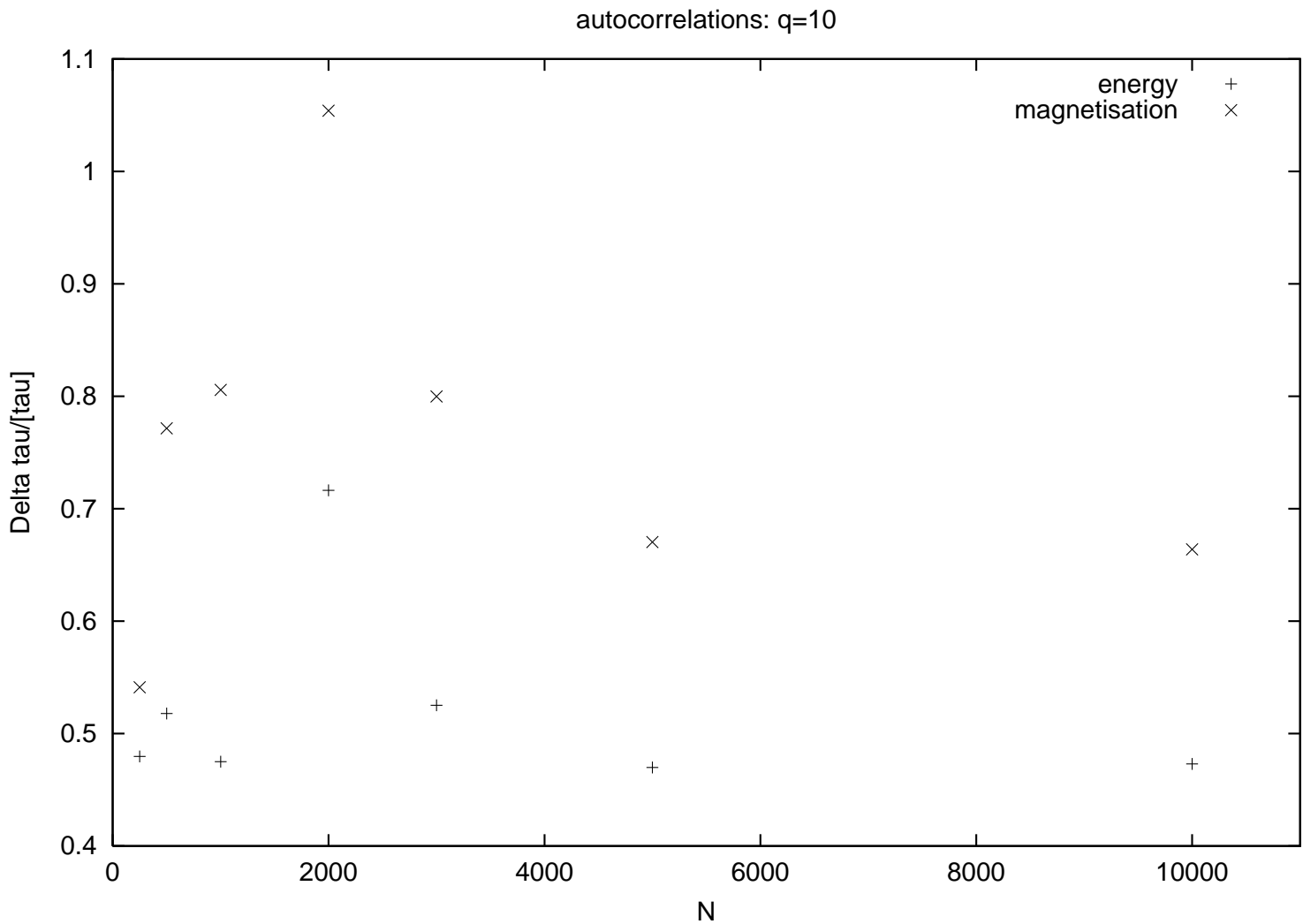}
\caption[]{\label{fig5} A plot of $\Delta \tau /  [ \tau ]_{av}$
for both the energy and magnetisation. Given the caveat in the text regarding
the simulation points this is clearly tending to
a constant for increasing system size.
}
\end{figure}                       
%-----------Figures------------------
\clearpage \newpage
\begin{figure}[htb]
\vskip 20.0truecm
\includegraphics{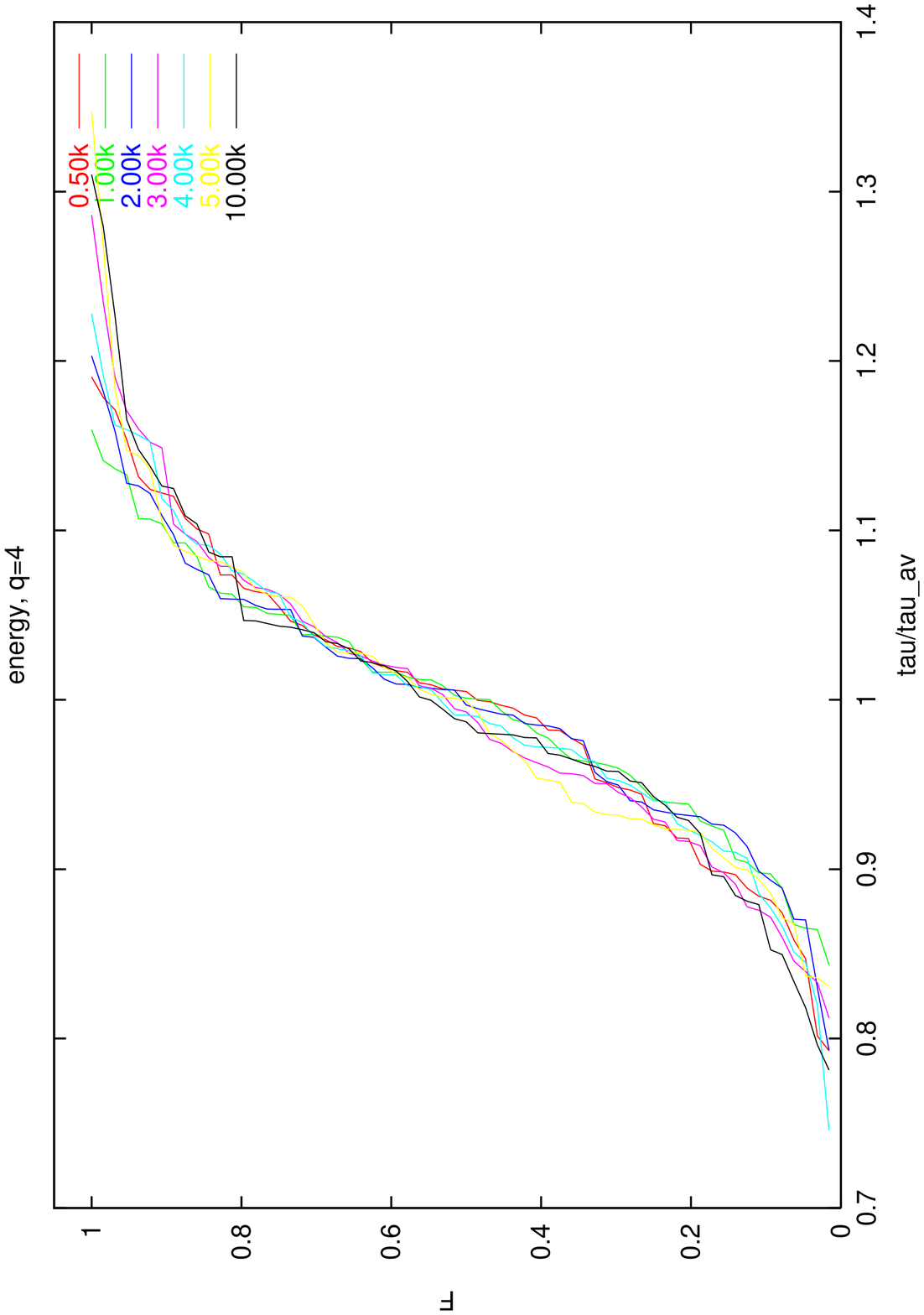}
\caption[]{\label{fig6} The scaled cumulative distribution function
for the energy autocorrelations in the $q=4$ Potts model showing
the good data collapse.
}
\end{figure}              
                      
%-----------Figures------------------
\clearpage \newpage
\begin{figure}[htb]
\vskip 20.0truecm
\includegraphics{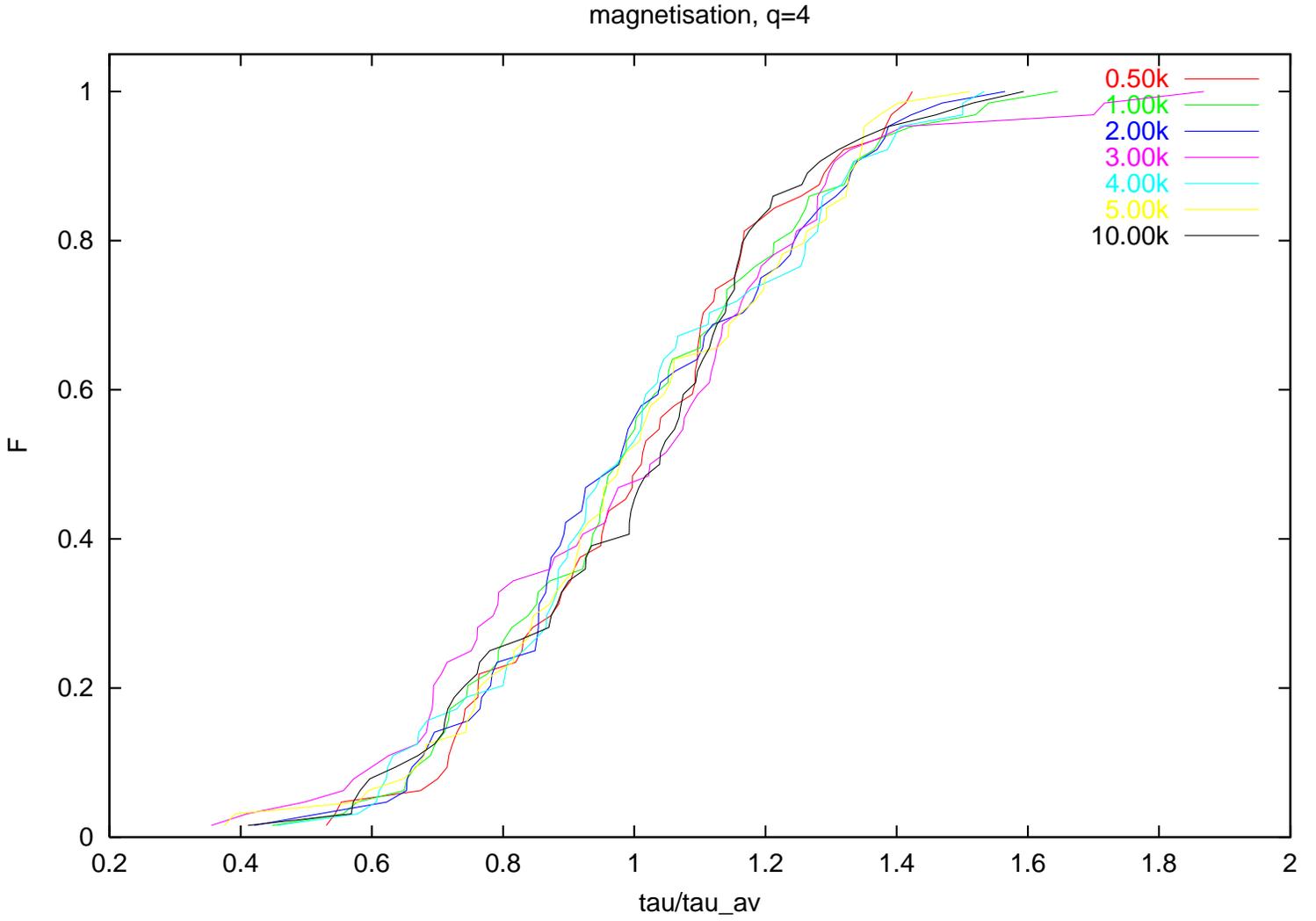}
\caption[]{\label{fig7} The scaled cumulative distribution function
for the magnetisation autocorrelations in the $q=4$ Potts model showing
a similar data collapse to the energy case.
}
\end{figure}       
%-----------Figures------------------
\begin{figure}[htb]
\vskip 20.0truecm
\includegraphics{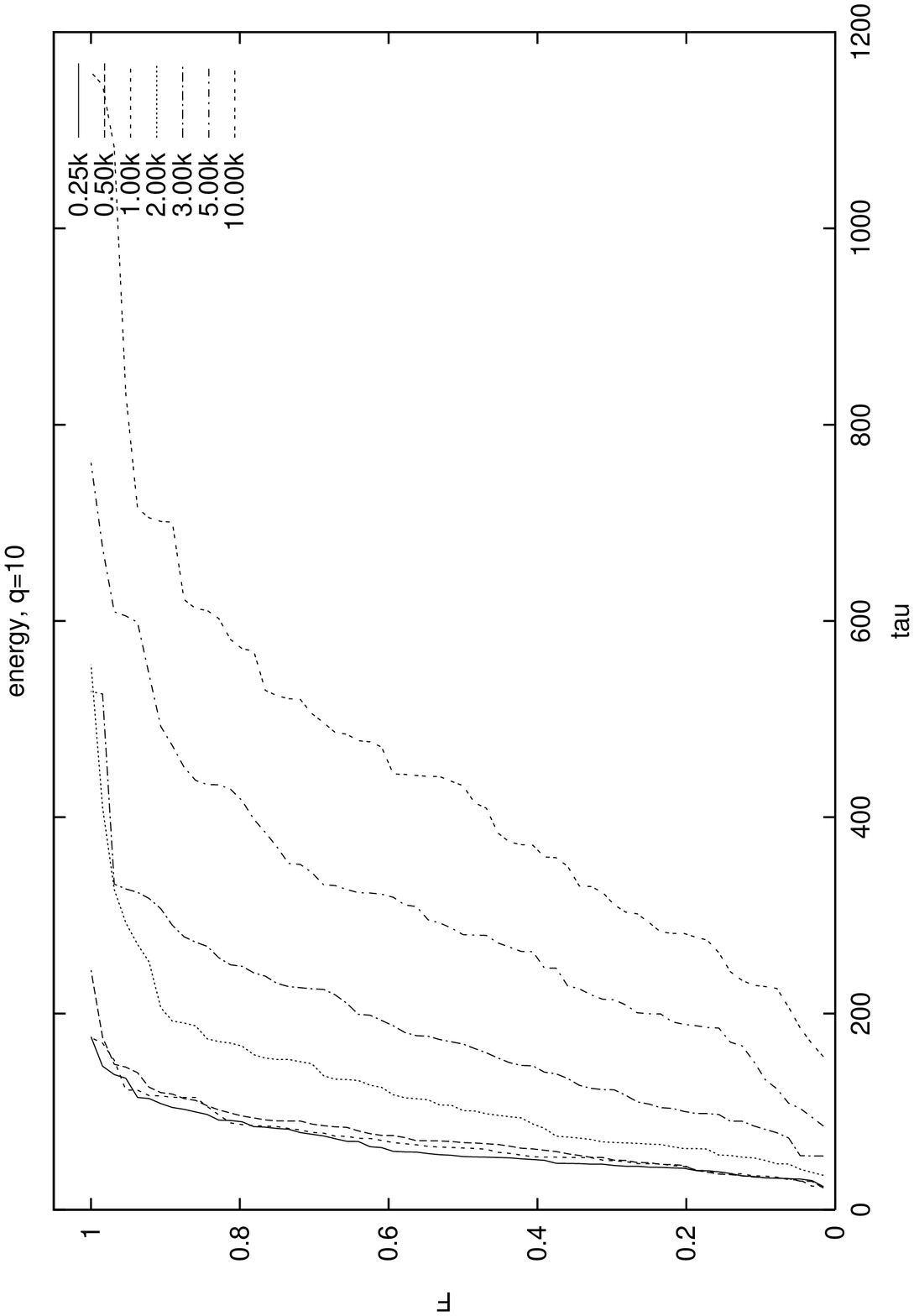}
\caption[]{\label{fig8a} The cumulative distribution function for the energy
autocorrelations in the $q=10$ Potts model.
The system size is {\it increasing} as
the curves move to the right.
}
\end{figure}
%-----------Figures------------------
\clearpage \newpage
\begin{figure}[htb]
\vskip 20.0truecm
\includegraphics{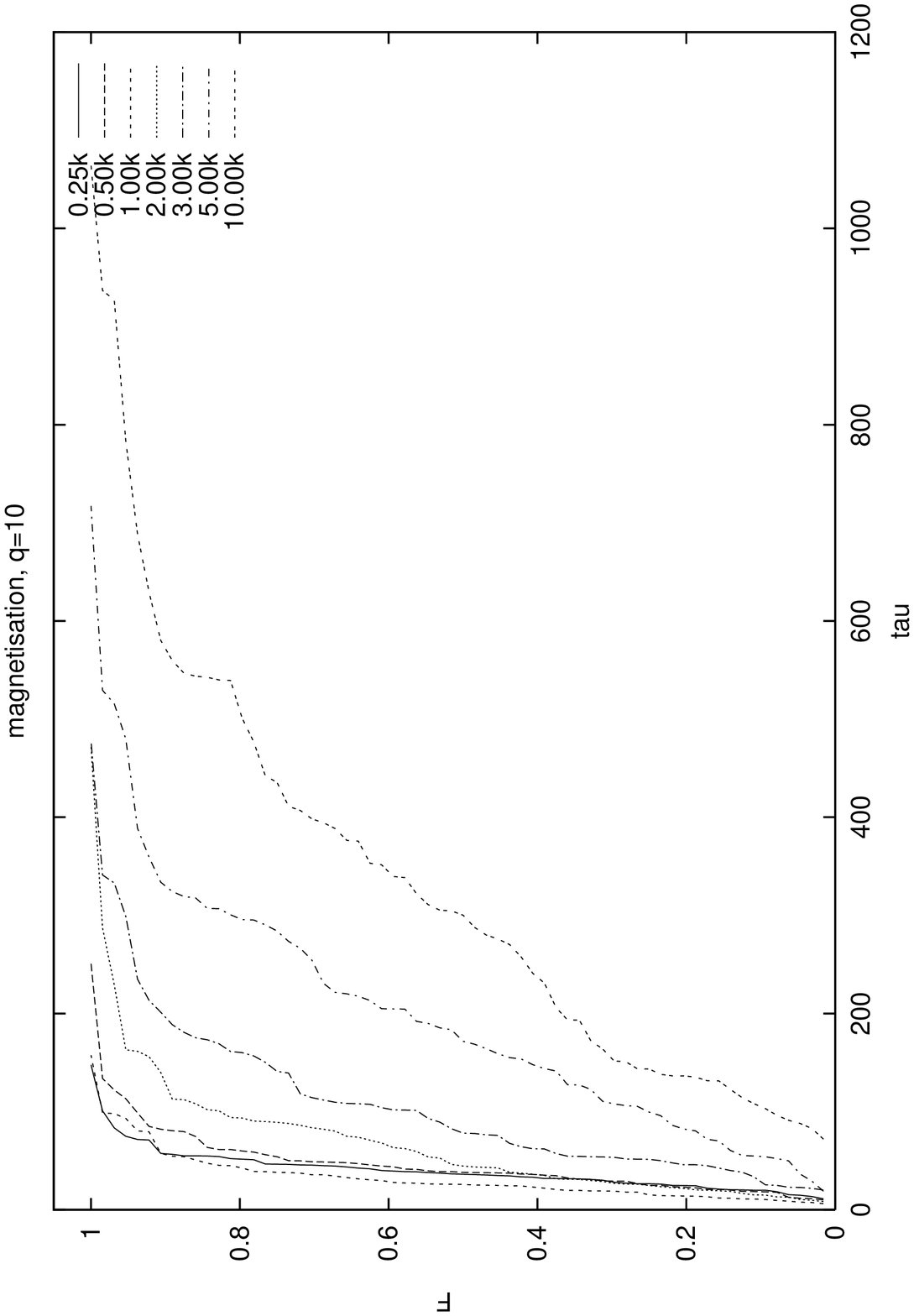}
\caption[]{\label{fig9a} The cumulative distribution function
for the magnetisation autocorrelations in the $q=10$ Potts model.
Again, the system size is increasing as
the curves move to the right.
}
\end{figure}
%-----------Figures------------------
\clearpage \newpage
\begin{figure}[htb]
\vskip 20.0truecm
\includegraphics{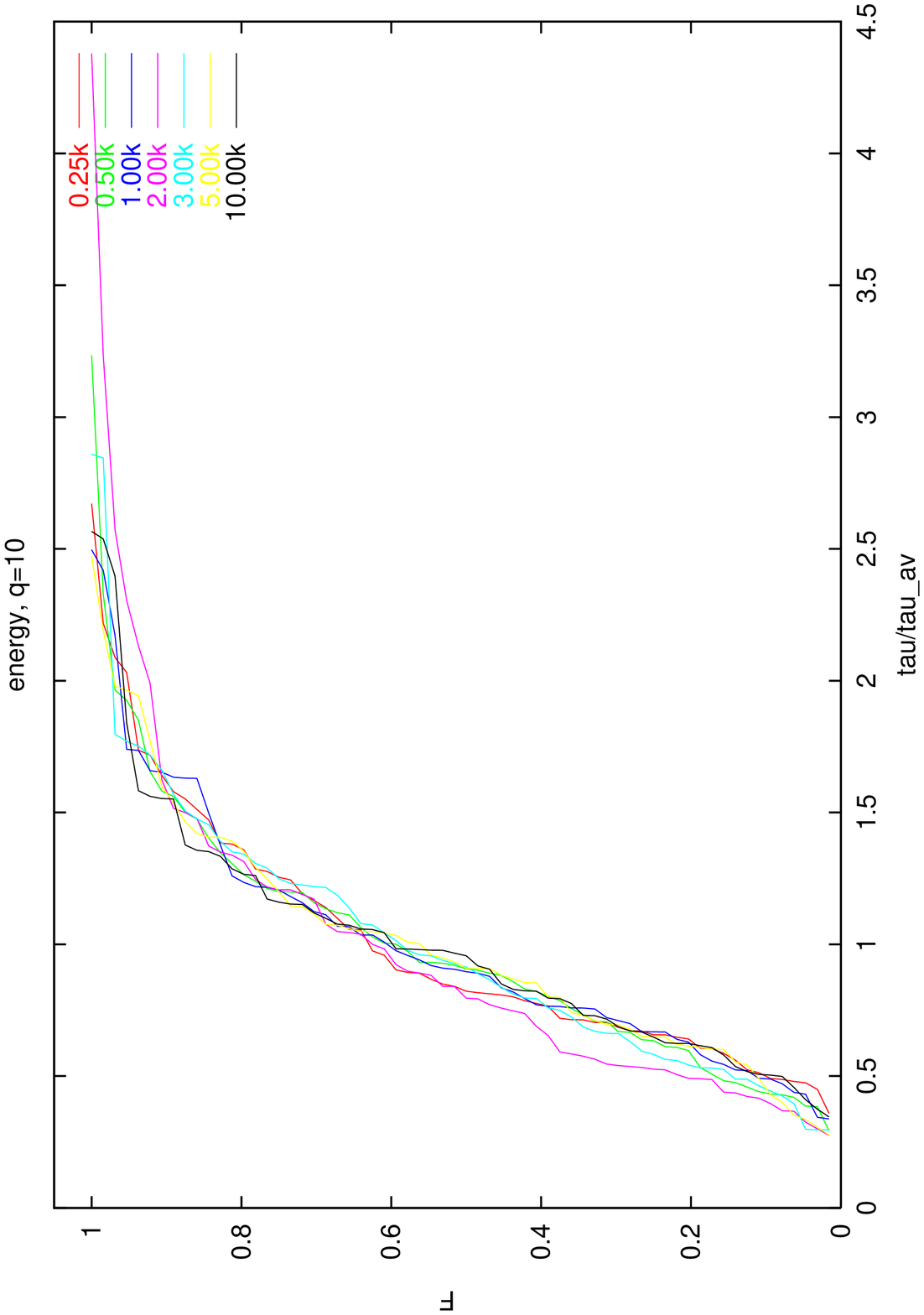}
\caption[]{\label{fig8} The scaled cumulative distribution function
for the energy autocorrelations in the $q=10$ Potts model showing
the data collapse.
}
\end{figure}      
%-----------Figures------------------
\clearpage \newpage
\begin{figure}[htb]
\vskip 20.0truecm
\includegraphics{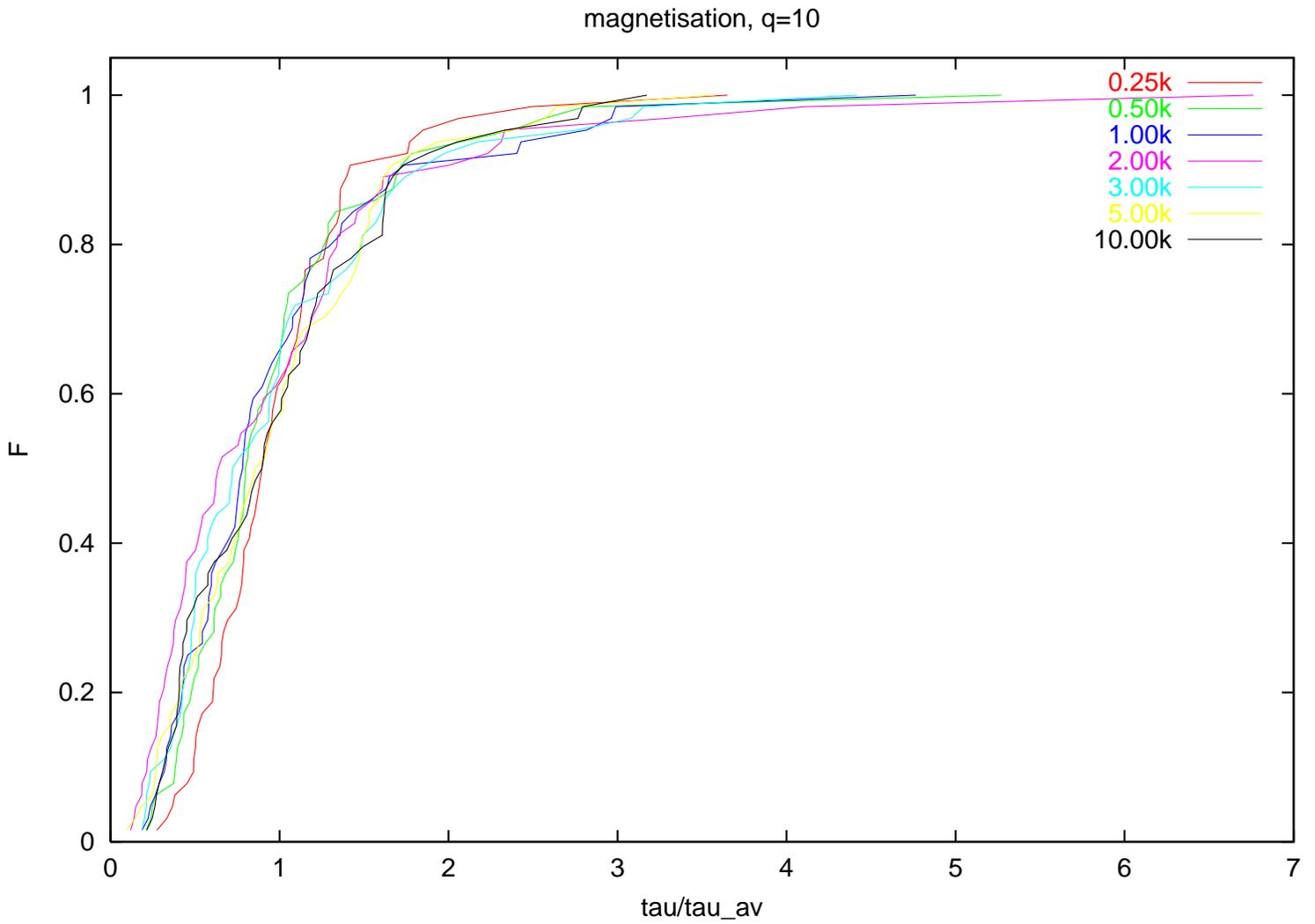}
\caption[]{\label{fig9} The scaled cumulative distribution function
for the magnetisation autocorrelations in the $q=10$ Potts model.
}
\end{figure}  
\end{document}